\documentclass[osajnl,twocolumn,showpacs,superscriptaddress,10pt]{revtex4-1} 
\usepackage{amsmath,amssymb,graphicx,color}
\usepackage{epstopdf}

\begin{document}

\title{Ultra-stable laser with average fractional frequency drift rate below $5\times10^{-19}/\mathrm{s}$}

\author{Christian Hagemann}
\author{Christian Grebing}
\author{Christian Lisdat}
\author{Stephan Falke}
\author{Thomas Legero}
\author{Uwe Sterr}\email{Corresponding author: uwe.sterr@ptb.de}
\author{Fritz Riehle}
\affiliation{
Physikalisch-Technische Bundesanstalt (PTB), Bundesallee 100, 38116 Braunschweig, Germany\\
$^*$Corresponding author: uwe.sterr@ptb.de
}

\author{Michael J. Martin}
\author{Jun Ye}
\affiliation{
JILA, National Institute of Standards and Technology and University of Colorado, Department of Physics, 440 UCB, Boulder, Colorado 80309, USA
}

\begin{abstract}
Cryogenic single-crystal optical cavities have the potential to provide highest dimensional stability. We have investigated the long-term performance of an ultra-stable laser system which is stabilized to a single-crystal silicon cavity operated at 124~K. 
Utilizing a frequency comb, the laser is compared to a hydrogen maser that is referenced to a primary caesium fountain standard and to the $^{87}\mathrm{Sr}$ optical lattice clock at PTB. 
With fractional frequency instabilities of 
$\sigma_y(\tau)\leq2\times10^{-16}$ 
for averaging times of 
$\tau=60\mathrm{~s}$ to $1000\mathrm{~s}$ and 
$\sigma_y(1~\mathrm{d})\leq 2\times10^{-15}$ 
the stability of this laser, without any aid from an atomic reference, surpasses the best microwave standards for short averaging times and is competitive with the best hydrogen masers for longer times of one day.
The comparison of modeled thermal response of the cavity with measured data indicates a fractional frequency drift below $5\times 10^{-19}$~/s, which we do not expect to be a fundamental limit.
\end{abstract}


\maketitle 

Advancements in ultra-stable oscillators push forward a wide range of cutting-edge experiments such as atomic clockwork \cite{san99,nic12,hin13}, or tests of fundamental physics \cite{eis09,cho10a}, and applications in deep space navigation \cite{gro10a} or very-long baseline interferometry \cite{nan11} at shorter wavelength \cite{doe11}. 
For these applications, both the short term instability on few second time scale and the long term stability over one day are essential.
So far, mostly frequency standards and oscillators operating in the microwave domain, such as hydrogen masers and cryogenic microwave oscillators for improved short term stability \cite{gro10a}, are used. 

Recently, oscillators operating in the optical domain have demonstrated fractional frequency instabilities with Allan deviation of $\sigma_y(\tau)= 1 \times10^{-16}$ at $\tau = 1 - 1000$~s \cite{nic12,mar13}. For most of the applications mentioned above it would be desirable if the short term stability of the oscillator could be kept also at much longer times without the need of a sophisticated optical atomic clock.  
However, aging in the commonly employed ultra-low expansion (ULE) glass cavities leads to a poorly predictable non-linear shrinkage of the resonator length $L$ on the order of 
$\dot{L}/L=10^{-17}/\mathrm{s}$ to $10^{-16}/\mathrm{s}$ 
\cite{dub09}, which translates to frequency drifts 
$\dot{\nu}/\nu=-\dot{L}/L$ of the cavity resonance. 
Due to residual temperature fluctuations of the cavity, the day-to-day predictability of the cavity drift rate remains at $\pm 2\times 10^{-17}/\mathrm{s}$ \cite{dub09}. 
With material creep being absent in single-crystals, crystalline reference cavities are a promising candidate for laser sources that keep their stability in the long-term. 
Already in 1998 dimensional stabilities of $7\times10^{-19}/\mathrm{s}$ have been achieved employing a single-crystal sapphire cavity at 4.3~K \cite{sto98} and short-term stability of $\sigma_y(20\mathrm{~s})=2.3\times10^{-15}$ \cite{see97}.
However, the automatic liquid-nitrogen refill processes led to fractional frequency excursions of $3\times10^{-13}$ every few hours.
\begin{figure}
\centerline{\includegraphics[width=\linewidth]{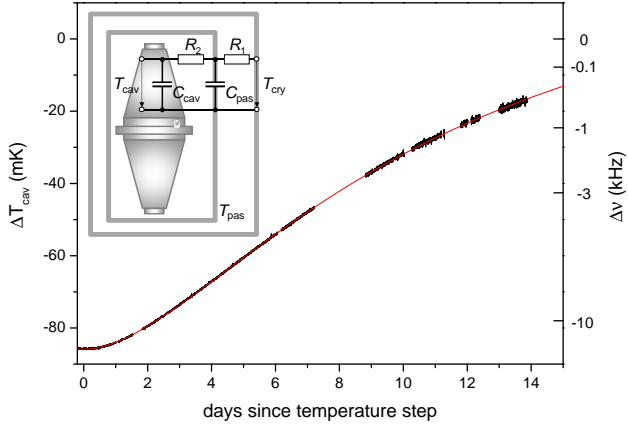}}
\caption{Response of silicon cavity temperature inferred from the laser frequency after the temperature $T_{\mathrm{cry}}$ of the cryogenic shield was increased by $+100$~mK (black points). The red trace shows a fit of the expected response function. The inset shows the setup with the employed thermal shields represented by an equivalent electric circuit of the thermal model.}
\label{fig:tc}
\end{figure}
\begin{figure}
\centerline{\includegraphics[width=\linewidth]{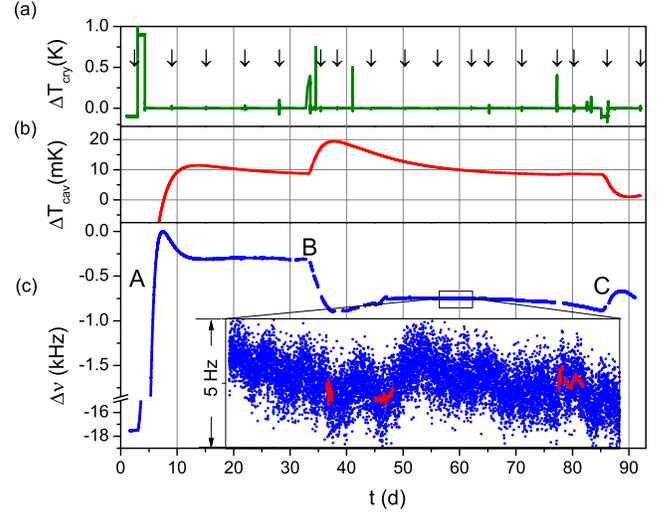}}
\caption{(a): Measured temperature deviation $\Delta T_\mathrm{cryo}$ of the cryogenic shield. The arrows indicate when the liquid nitrogen storage Dewar of the cryostat was refilled. 
(b): Deviation $\Delta T_\mathrm{cav}$ of the cavity from zero-crossing temperature.
(c): Absolute frequency of the laser stabilized to the silicon cavity with respect to the zero-crossing frequency $\nu(T_0)$ at $t=8~$d. The frequency was referenced to a hydrogen maser (moving average, window: $1000$~s) which was referenced to the Cs fountain clock.
The inset shows a zoom of $\Delta \nu$ between days 56.5 and 63 including frequency data derived from comparison with the $^{87}$Sr optical lattice clock (red dots).}
\label{fig:freq}
\end{figure}
Recently, we have set up a laser system based on a single-crystal cryogenic silicon cavity. 
Silicon possesses a zero-crossing in the Coefficient of Thermal Expansion (CTE) at a relatively high temperature of $T_0=124.2\mathrm{~K}$, which greatly reduces the technical demands of cooling. 
This laser system has demonstrated a short-term stability of the modified Allan deviation \cite{all81}
$\mathit{Mod}\;\sigma_y (\tau) \leq 1\times10^{-16}$ 
for averaging times of 
$\tau=0.1\mathrm{~s}$ to $1\mathrm{~s}$ \cite{kes12a}. 
The crystalline nature also lets us expect high long-term stability \cite{kes12a}. 
Here we present our measurements with this system obtained over an extended period of time, where the frequency of the laser system has been compared to the $^{87}$Sr optical lattice clock at PTB \cite{fal11,fal13} and to a hydrogen maser which is referenced to a primary Cs fountain clock \cite{ger10}. 

To minimize the effect of any remaining drift of the cavity temperature $T_\mathrm{cav}$ on its length, operating the cavity as close as possible to the zero CTE temperature $T_0$ is crucial. 
For small deviations $\Delta T_{\mathrm{cav}} = T_{\mathrm{cav}}-T_0$ the length and frequency of the silicon cavity depend on its temperature 
$T_{\mathrm{cav}}$  according to
\begin{equation}
\label{cte}
-\frac{\Delta \nu(T_\mathrm{cav})}{\nu(T_0)}
=\frac{\Delta L}{L(T_0)}=\frac{1}{2} \alpha'(T_0) \Delta T_\mathrm{cav}^2 \;,
\end{equation}
where 
$L(T_0) \approx 21\mathrm{~cm}$ is the minimum length of the cavity at $T_0=124.2$~K,  
and 
$\alpha'=1.7\times10^{-8}~\mathrm{K}^{-2}$ 
is the slope of the CTE around $T_0$ \cite{kes12a}. 
A low-vibration cryostat employing cold nitrogen gas has been developed which provides a cryogenic actively controlled shield at $T_\mathrm{cry}$ with sub-millikelvin temperature stability. 
An additional passive shield further dampens residual temperature fluctuations. 
Adjusting the cryostat temperature to $T_{\mathrm{cav}}=T_0$ is quite cumbersome in the present setup, as no temperature sensor is attached to the silicon cavity to best preserve its vibration-insensitive mounting design and avoid local heating. 
Only the temperature $T_\mathrm{cry}$ at the bottom of the actively controlled shield was measured. 
As there are temperature gradients across that shield, a cavity temperature $T_0$ corresponds to a temperature $T_\mathrm{cry}$ about $0.5~$K below $T_0$, which is taken into account in our analysis.
With knowledge of $\nu(T_0)$ and assuming that $\nu(T_0)$ is highly constant, the cavity temperature $T_{\mathrm{cav}}$ can be estimated using Eq.~(\ref{cte}) from the observed absolute laser frequency $\nu$. 
Following this approach, we can use the response of the cavity frequency to a $+100$~mK step of $T_\mathrm{cry}$ to determine the thermal dynamics of the system (Fig.~\ref{fig:tc}), which behaves as a second order thermal low-pass:
\begin{align}
\label{eq:temp_time}
\dot T_\mathrm{pas}& 
= \frac{T_\mathrm{cry}-T_\mathrm{pas}}{\tau_1}-\frac{T_\mathrm{pas}-T_\mathrm{cav}}{\tau_2}\;,\\
\label{eq:temp_time2}
\dot T_\mathrm{cav}& 
= \frac{T_\mathrm{pas}-T_\mathrm{cav}}{\tau_3}\;
\end{align}
with time constants 
$\tau_1=R_1 C_\mathrm{pas}$, $\tau_2=R_2 C_\mathrm{pas}$ and $\tau_3=R_2 C_\mathrm{cav}$. 
From the thermal response of the cavity to a temperature step on the cryogenic shield (red curve in Fig.~1) and the known heat capacities 
$C_\mathrm{pas}\approx 3\mathrm{~kJ/K}$, 
$C_\mathrm{cav}\approx 0.5\mathrm{~kJ/K}$,
we find the thermal resistivities 
$R_1\approx 60\mathrm{~K/W}$, 
$R_2\approx 1850\mathrm{~K/W}$,
and the time constants
$\tau_1=2.1$~d, $\tau_2=64.2$~d, and $\tau_3=10.7$~d. 
The thermal model allows to calculate the cavity temperature $T_\mathrm{cav}$ from the measured $T_\mathrm{cry}$.

With the knowledge of the dynamics of the thermal system we can rapidly tune the silicon cavity to $T_0$, which we did at the beginning of a frequency comparison of three months duration (Fig.\ref{fig:freq}). From an initial laser frequency measurement, temperature offsets of 
$\Delta T_\mathrm{cav} \approx-100$~mK and 
$T_\mathrm{pas}-T_\mathrm{cav} \approx1$~mK were determined. 
To quickly increase $T_\mathrm{cav}$, $T_\mathrm{cry}$ was rapidly raised by 1~K and subsequently lowered by 0.9~K once $T_\mathrm{cav}$ has increased by 33~mK (mark A). 
$T_\mathrm{cav}$ slightly overshot its zero-crossing temperature, 
settling at $\Delta T_\mathrm{cav}\approx 8~$mK within twenty days (see Fig.\ref{fig:freq}(b)). 
At mark B a failure of the cryostat's temperature control interrupted the thermalization of the system leading to an increased temperature $\Delta T_\mathrm{cav}\approx$ 18~mK within the next days. 
Shortly before the frequency comb became unavailable due to maintenance, a deliberate temperature decrease of the cryo-shield by $-100$~mK (mark C) over one day was introduced to better approach $T_0$. 
During the measurement period the temperature response of the silicon cavity was recorded via absolute frequency measurement of the laser stabilized to the cavity (see Fig.\ref{fig:freq}(c)).

\begin{figure}
\centerline{\includegraphics[width=0.9\linewidth]{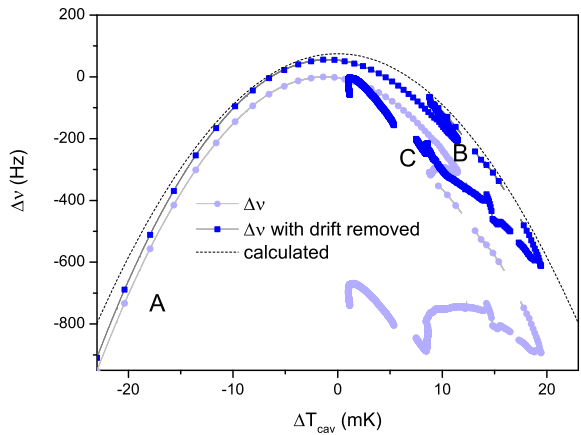}}
\caption{
Measured cavity frequency $\Delta \nu$ vs. calculated cavity temperature offset $\Delta T_\mathrm{cav}$ from the CTE crossing  (light blue symbols) using the temporal data shown in Fig \ref{fig:freq} that start near label A. To describe the behavior expected from the silicon thermal expansion (dashed line), a constant drift of $-8~$Hz/d ($-90~\mu$Hz/s) had to be assumed (drift-removed data shown as dark blue symbols).
The characters refer to the temperature perturbations indicated in Fig. \ref{fig:freq}. 
}
\label{fig:FreqvsT}
\end{figure}

From these data we have estimated the stability of the system: 
Between the two intentional changes of the cryo-shield temperature between mark A and C, over 70 days the laser frequency remained within 600~Hz, which would correspond to an average fractional frequency drift of $5\times10^{-19}/\mathrm{s}$. 
%
A more thorough investigation confirms this rough estimate: When plotting the laser frequency as a function of the calculated cavity temperature $T_\mathrm{cav}$ (Fig.\ref{fig:FreqvsT}) the curve shows the expected parabolic dependency once a constant frequency drift of $-8$~Hz/d ($\dot{\nu}/\nu=-4.8\times10^{-19}$/s) is corrected in the data. Possible causes for the residual negative drift corresponding to a change of optical length of $8~$fm/d could be isothermal creep of the resonator mirror coatings or buildup of adsorbed layers from the residual gas on the mirror surfaces. 
The minor deviation of an exact parabolic shape may be due to mechanical relaxations of the cavity mount as a result of the cryostat failure. 
The lowest drift was observed between day 56 and day 63 (see inset of Fig.\ref{fig:freq}(c)), where the average frequency drift of the laser was $1.7$~$\mu$Hz/s ($\dot{\nu}/\nu=9\times10^{-21}/\mathrm{s}$) 
due to compensation of drift and temperature variation.
The frequency instability compared to the hydrogen maser (corrected for its drift by comparison to the Cs fountain clock) is shown in Fig.~4. Similar results are obtained after removal of a linear cavity drift from all other data sets, where the temperature was unperturbed. 
Up to $\tau=10^4$~s, the instability curve follows the typical performance of such type of maser (dashed line) while the moderate increase in frequency instability from $\tau=10^4$~s to $10^5$~s indicates a contribution from the laser system.
\begin{figure}
\centerline{\includegraphics[width=0.9\linewidth]{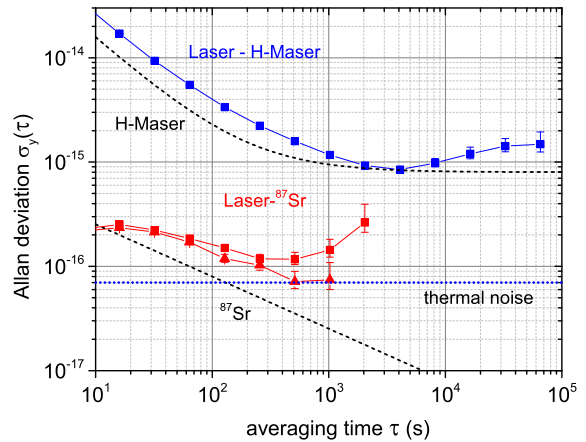}}
\caption{
Fractional frequency instability of the frequency data shown in the inset of Fig. \ref{fig:freq}. The red curves show the Allan deviations of the two longest continuous data sets from the comparison to the $^{87}$Sr clock (during days 59 and 62). The black dashed lines show the expected instabilities of a typical hydrogen maser and the $^{87}$Sr clock. The blue dotted line shows the Allan deviation expected from thermal noise.}
\label{fig:adev}
\end{figure}
The laser instability for shorter averaging times was inferred from a frequency comparison with a $^{87}\mathrm{Sr}$ lattice clock which was operational for several times in this time interval (inset of Fig.\ref{fig:freq}(c) red traces). The correlation with frequency excursions observed with the maser indicates that the instability observed in Fig.~4 above $10^4$~s is originating from the Si-cavity laser. 
During these measurements the lattice clock instability was $\sigma_y(\tau)=8\times10^{-16}/\sqrt{\tau/s}$;
about a factor of two bigger than its optimal stability \cite{hag13}, as the duty cycle of the clock was reduced to lower the systematic uncertainties during a frequency comparison \cite{fal13}. 
The observed instability of the Si-cavity laser system is below $\sigma_y=2\times10^{-16}$ for averaging times longer than one minute and reaches the $10^{-16}$-level at averaging times above $\tau=500$~s. 
After that the instability varies depending on the actual frequency fluctuations of the laser that occur at the time the strontium clock was available for comparison. 

The increase of laser instability between $\tau=10^3$~s and $10^5$~s is not yet understood and is not a property of the maser. 
Although Residual Amplitude Modulation (RAM) from the fiber-coupled phase modulator is actively canceled in the setup \cite{kes12a}, there might be uncompensated RAM and frequency pulling of the cavity resonance from parasitic reflections. 
As the finesse of the TEM$_{00}$-mode was poor (80~000), we employed the TEM$_{01}$-mode for cavity-locking \cite{kes12}. 
Thus parasitic effects from beam pointing cannot be ruled out as such effects had been observed previously. 
Also temperature fluctuations during a day would show up at this time scale. 

In conclusion, with fractional frequency instabilities of 
$\sigma_y(\tau)\leq2\times10^{-16}$ 
for averaging times of 
$\tau=60\mathrm{~s}$ to $1000\mathrm{~s}$ and 
$\sigma_y(1~\mathrm{d})\leq 2\times10^{-15}$ 
the stability of this laser surpasses the best microwave standards for short averaging times.
With an average daily drift of $4\times10^{-14}$ estimated from the two frequency values at the CTE zero-crossing after points A and C, it outperforms the best cryogenic sapphire microwave oscillators showing  
$-8.5\times 10^{-14}$/d \cite{tob06} and its stability is also competitive with the best hydrogen masers for times of up to one day.

The demonstrated frequency stability of a silicon single-crystal stabilized laser in the different regimes of averaging times does not represent a fundamental limit.   
In a second silicon cavity system that is currently under construction, a higher finesse cavity will allow a corresponding reduction of the influence of RAM and other parasitic effects that might result from the employment of the TEM$_{01}$-mode for cavity-locking. It has been shown recently that the influence of the RAM on the frequency stability can be further reduced to a fraction of $1 \times 10^{-6}$  of a cavity linewidth \cite{zha14}. 
This second cavity can be operated with reduced laser power, with less heating of the cavity from the absorbed laser power that currently amounts to $50~$mK from $25~\mu$W absorbed power. 
An optimized distribution of improved temperature sensors will enable a more careful adjustment and monitoring of the cavity temperature.
 
With these optimizations it seems feasible to realize an ultra-stable oscillator that provides fractional frequency stabilities below $\sigma_y(\tau)=1\times 10^{-16}$ from a second to a day. 
For example, such an ultra-stable oscillator would be useful in the comparison between optical frequency standards and primary Cs clocks, that requires averaging times of several days to reach their uncertainty of $\Delta\nu/\nu\approx10^{-16}$. 
With the improvements now under way it will also be possible to elucidate the nature of the observed temporal drift rate of about $-8$~Hz/d. Its constant value observed up to now would furthermore allow to apply a feed forward correction loop to achieve an even higher long-term stability. In the region of several days around day 60 (Fig. 2 c) where the drift rate is compensated by another -- presumably temperature -- effect a fractional instability of $2 \times 10^{-19}$ was observed.
As optical clocks still lack sufficient reliability for unattended long-term operation, the very small observed long-term drift rate will allow such a laser to act as a flywheel oscillator that could bridge longer gaps in measurement and which would facilitate comparisons between optical clocks and long-term measurement with microwave time standards. 
\section*{Acknowledgments}
We thank A. Bauch and S. Weyers for providing the frequency data between Cs~fountains and H~masers. 
The cryogenic silicon cavity laser system employed in this work was developed jointly by the JILA Physics Frontier Center (NSF) and the National Institute of Standards and Technology (NIST), the Centre for Quantum Engineering and Space-Time Research (QUEST) and Physikalisch-Technische Bundesanstalt (PTB). 
We acknowledge funding from the DARPA QuASAR program, the European Community ERA-NET-Plus Programme (Grant No. 217257), the European Community 7$^\mathrm{th}$ Framework Programme (Grant Nos. 263500) and the European Metrology Research Programme (EMRP) under IND14. 
The EMRP is jointly funded by the EMRP participating countries within EURAMET and the European Union.
J.~Y.~acknowledges support from the Humboldt foundation.

\end{document}